\documentclass[a4paper,12pt]{article}
\usepackage{epsfig}
\usepackage{amssymb}
\begin{document}
\thispagestyle{empty}

\newcommand{\etal}  {{\it{et al.}}}  
\def\Journal#1#2#3#4{{#1} {\bf #2}, #3 (#4)}
\def\PRD{Phys.\ Rev.\ D}
\def\NIMA{Nucl.\ Instrum.\ Methods A}
\def\PRL{Phys.\ Rev.\ Lett.\ }
\def\PLB{Phys.\ Lett.\ B}
\def\EPJ{Eur.\ Phys.\ J}
\def\IEEETNS{IEEE Trans.\ Nucl.\ Sci.\ }
\def\CPCD{Comput.\ Phys.\ Commun.\ }



{\Large\bf
\begin{center}
Dilaton decays into unparticles and a single photon
\end{center}
}

\begin{center}
\large{ G.A. Kozlov  }
\end{center}
\begin{center}
 { Bogolyubov Laboratory of Theoretical Physics\\
 Joint Institute for Nuclear Research,\\
 Joliot Curie st., 6, Dubna, Moscow region, 141980 Russia }
\end{center}
\begin{center}
 \large{I.N. Gorbunov }
\end{center}
\begin{center}
 { University center\\
 Joint Institute for Nuclear Research,\\
 Joliot Curie st., 6, Dubna, Moscow region, 141980 Russia }
\end{center}

 \begin{abstract}
 \noindent
 {We study the production of the vector $U$ -unparticle stuff  and a single photon in decays of a dilaton.
The signals of an unparticle can be detected through the missing energy and momentum
distribution carried away by  $U$ once it was produced in radiative decay of a dilaton The continuous energy spectrum of the emitted photons encoding the recoil unparticle can be measured in precision studies of rare decays of the dilaton or Higgs-boson after their discoveres.}


\end {abstract}
PACS numbers: 12.60.Cn, 12.60.Fr, 14.80.Va, 11.25.Hf




\bigskip

{\it Introduction.-}  There is a class of models in which the electroweak symmetry breaking (EWSB) at the
scale $v\simeq$ 246 GeV (the vacuum expectation value of the $SU(2)$ doublet Higgs scalar field) can, for example,  be triggered by spontaneous breaking of the scale  symmetry at an  energy scale $f\geq v$ [1,2]. In this scenario, one may have nearly conformal dynamics at a scale
$ \sim 4\pi f$ below which the scale symmetry is broken and one feeds into an electroweak (EW) sector.

The spectrum may contain an EW singlet scalar field $\chi (x)$, the dilaton mode, that is the
pseudo-Goldstone boson associated with the spontaneous breaking of conformal symmetry  and this dilaton may have Higgs-like properties.
The real dilaton field $\sigma (x)$ is parametrized by $\sigma (x) = \chi (x) - f$ with the order parameter $\langle \chi (x)\rangle = f$, the symmetry breaking scale. The Higgs-boson can be seen as a dilaton in the limit case $ f = v$.  The mass of the $\sigma$-dilaton is naturally light, $m_{\sigma} \sim \epsilon\cdot f$, where the small parameter $\epsilon$ controls the deviations from exact scale invariance. The dilaton becomes massless when the conformal invariance is recovered.

Since of its pseudo-Goldstone nature, the dilaton could be the  messenger field between
the Standard Model (SM) fields and the hidden sector. For example, the dilaton field itself can be lighter than, e.g., the dark matter (DM) particle, and be the dominant product of DM annihilation.
Note that if one assumes that both SM and DM are fully embedded in the conformal
sector, one can propose that the dilaton is the dominant messenger between the DM, the SM and
the unparticle stuff. The latter itself with the scaling dimension $d$ may appear as a non-integer
number $d$ of invisible particles [3].
We also assume that the dilaton could be the dominant origin of the unparticle production through the SM fields. A set of phenomenological implications of the unparticle stuff can be found in [4].  The results of direct searches for the Higgs-boson at LEP and the Tevatron adapted to constrain both the dilaton mass $m_{\sigma}$ and the conformal breaking scale $f$ has been carried out in [5].  In particular, based on the Tevatron data, there is a widely allowed range for a light dilaton below 200 GeV even for $f \sim  O(1)$ TeV.

In this letter, we present the model in which  the decay of a dilaton into a vector unparticle $U$ and a single photon, $\sigma\rightarrow\gamma\,U$, is studied. It is related to the models introduced by Ferrari [6] in study of the Higgs phenomenon and by Zwanziger [7] in the lesson from a soluble model in quantum electrodynamics from which the present model is distinguished by the presence of the nearly conformal sector and the couplings to the unparticle stuff.
Signals of the latter can be detected in the missing energy and momentum distribution carried away by  the unparticle once it was produced in decay of a dilaton. The coupling of a dilaton to unparticle stuff is through the loop composed with the  quark fields flowing in the loop. The attractive feature of the decay  $\sigma\rightarrow\gamma\,U$ is that the photon energy has a continuous spectrum in the rest frame of $\sigma$, in contrast to, e.g., $\sigma\rightarrow\gamma\gamma$ or $\sigma\rightarrow\gamma\,Z$. In addition, this mode would predict an instructive tool for the study of the dilaton in respective broad range of the dilaton mass up to the order $O(1)$ TeV, and the distiguishing the dilaton from the SM Higgs-boson $H$, restricted by its mass with 130-140 GeV in the decays like $H\rightarrow\gamma\,U$, $H\rightarrow\gamma\gamma$, $H\rightarrow\gamma\,Z$, etc.

{\it Setup.-} We suppose the production of the dilaton through the gluon-gluon fusion, $g\,g\rightarrow \sigma$.
The dilaton coupling to quarks induces the  couplings, $\sigma g\,g$, which are sensitive to heavy quarks present in any extension of the SM and the contribution of which are embedded in the beta-function. Since the couplings $\sigma g\,g$ are crucial for collider phenomenology, it has been shown [2] that these couplings can be significantly enhanced under very mild assumption about high scale physics. At energies below the scale $4\pi f$ the effective dilaton couplings to massless gauge bosons are provided by the SM quarks  lighter than the dilaton:
$\sigma [c_{EM} (F_{\mu\nu})^{2} + c_{s} (G^{a}_{\mu\nu})^{2}]/(8\,\pi\,f) $. Here, $F_{\mu\nu}$ and
$G^{a}_{\mu\nu}$ are the electromagnetic (EM) and gluon fields strength tensors, respectively;
$c_{EM} = - \alpha\cdot 17/9$ if $m_{W} < m_{\sigma} < m_{t}$, $c_{EM} = - \alpha\cdot 11/3$
if $m_{\sigma} > m_{t}$;
$c_{s} = \alpha_{s}\cdot (11 - 2\,n_{light}/3)$; $n_{light}$ is the number of quarks lighter than the dilaton;
$\alpha$ and $\alpha_{s}$ are EM and strong coupling constants, respectively. The second term
in the effective coupling above mentioned indicates a $(33/2 - n_{light})$-factor increase of the
coupling strength compared to that of the SM Higgs boson.

The upper limit of $f$ can be estimated using the statistical significance of the Higgs signal
in experiment. Assuming the dominant production of the dilaton via the gluon fusion process,
the significance of the dilaton signal is given by the following rescaling [2]
$$\left (\frac{S}{\sqrt {B}}\right )_{\sigma} = \frac{c^{2}_{s}}{\alpha^{2}_{s}}\,\frac{v^{2}}
{f^{2}}\,\left (\frac{S}{\sqrt {B}}\right )_{Higgs}.$$
Based on the statistical significance of the Higgs signal, e.g., at ATLAS [8,9]  when the product $(c^{2}_{s}/\alpha^{2}_{s})\,(v^{2}/f^{2}) > 1/8$  at 100 $fb^{-1}$ integrated luminosity   for $m_{\sigma} > 160 $ GeV, one can find: $f < 5.33$ TeV if the dilaton is lighter than the top quark, or $f < 4.87$ TeV otherwise.

{\it Model.-}
The model is formulated in terms of a Lagrangian which features the dilaton field $ \sigma (x)$ as the local one and from which the vector potential $A_{\mu} (x)$ is derived, the conformal field given by the operator $O_{U}^{\mu}$ and a set of the SM fields. The conformal invariance can be broken by the couplings with non-zero mass dimension effects.
The Lagrangian density (LD) with a small explicit breaking of the conformal symmetry is $ L = L_{1} + L_{2}$, where
\begin{equation}
\label{e1}
L_{1} = -B\partial_{\mu}A^{\mu} + \frac{1}{2\xi}B^{2} -\frac{1}{\Lambda_{U}^{d-3}}(A_{\mu}
- \partial_{\mu} \sigma)O^{\mu}_{U} + \bar\psi (i\hat {\partial} - m +g\hat {A})\psi
 -  \frac {\sigma}{f}\sum_{\psi} \left (m + \epsilon y_{\psi}v\right ) \bar\psi\psi  ,
\end{equation}
\begin{equation}
\label{e2}
L_{2} =
 \frac{1}{\Lambda_{U}^{d-1}}\left [\sum_{q} \bar \psi (c_{v}\,\gamma^{\mu} -  a_{v}\,\gamma^{\mu}\gamma_{5})
\psi\, O_{{U}_{\mu}} + \frac{1}{\Lambda^{2}_{U}} W^{a}_{\mu\alpha}W^{a\mu}_{\beta}\left (\partial^{\alpha}
O^{\beta}_{U} + \partial^{\beta}O^{\alpha}_{U}\right )\right ] .
\end{equation}

The dilaton acquires a mass and its couplings to quarks can undergo variations from the standard form. In particular, since scale symmetry is violated by operators involving quarks, shifts in the dilaton Yukawa couplings to quarks can appear. This is given by $\epsilon = m^{2}_{\sigma}/f^{2}$ which parametrizes the size of the deviation from exact scale invariance [10]. In LD (\ref{e1}) the nine additional contributions to Yukawa couplings $y_{\psi}$ are taken into account ($y_{\psi}$ are $3\times 3$ diagonal matrices in the flavor space);   $\psi (x)$ stands for the spinor field with the mass $m$; the real parameter $\xi > 0$ and $g$ is the coupling constant.  The field $B(x)$ plays the role of the gauge-fixing multiplier, and it remains free. The vector operator $O_{U}^{\mu}$ describes a scale-invariant hidden sector that possess an infra-red fixed point at a high scale $\Lambda_{U}$, presumably above the EW scale. $c_{v}$ and $a_{v}$ are vector and axial-vector couplings. In the case of a Higgs-boson decays into two photons, the $W$-bosons contribute more significantly. In the model considered here, the only  SM quarks contribution is saved, because the $W$-boson loop contribution is suppressed by two more powers of $\Lambda_{U}$ in (\ref{e2}), and due to significantly large value of $\Lambda_{U}$ one can ignore it.

The equations of motion  are ($\nabla \equiv \partial /\partial x_{\mu}$)
\begin{equation}
\label{e3}
\partial_{\mu} \sigma \simeq A_{\mu} -  \frac{1}{\Lambda_{U}^{2}} \bar \psi (c_{v}\,\gamma_{\mu} -  a_{v}\,\gamma_{\mu}\gamma_{5}) \psi,
\end{equation}

$$\partial_{\mu}\,A^{\mu} = \xi^{-1}\,B, $$

$$\partial_{\mu} B = - J_{\mu} + \frac{1}{\Lambda_{U}^{d-3}}\,O_{U_{\mu}}, \,\,\,
J_{\mu} = g\,\bar{\psi}\,\gamma_{\mu}\,\psi,$$

$$\frac{1}{\Lambda^{d-3}_{U}}\,\partial_{\mu}O_{U}^{\mu} +  \frac{1}{f} (m + \epsilon\,y_{\psi}\,v)
\bar {\psi}\,\psi = 0, $$

$$\left [i\,\hat {\partial} - m\,\left (1+\frac{\sigma}{f}\right ) + g\,\hat {A} - \frac{\sigma}{f}\,
\epsilon\,y_{\psi}\,v + \frac{1}{\Lambda_{U}^{d-1}}\, O^{\mu}_{U} (c_{v}\,\gamma_{\mu} -  a_{v}\,\gamma_{\mu}\gamma_{5})\right ]\psi = 0. $$
We assume $\xi\neq\infty$  in (\ref{e1}) since otherwise the model becomes trivial.

In the nearly conformal sector the equation (\ref {e3}) looks like the dipole equation
$\left (\nabla^{2} \right )^{2} \sigma (x) \simeq 0$, where $\sigma (x)$ is the dipole ghost-like field. It is supported by the weakly changing operator $O^{\mu}_{U}$ in space-time and the conservation of the current $J_{\mu}$.
In this case, the following commutators
$$[B(x), \sigma (y)] = -i\,D(x-y),\,\,[B(x), B(y)] = 0, \,\, [\sigma (x), \sigma (y) ] = - i\,\xi^{-1}\, F(x-y), $$
are evident  for all $x$ and $y$, where $D(x) = (2\,\pi)^{-1}\,\delta (x^{2}) sgn (x^0)$ is the Pauli-Jordan function with the properties $\nabla^{2} D(x) =0$, $D(0,\vec{x}) =0$,
$\partial_{0} D(0,\vec{x}) = \delta^{3}(\vec{x})$; $F(x) = (8\,\pi)^{-1}\,\theta (x^{2})sgn (x^0)$.

The propagator $\tilde W(p)$ of the ghost-like dilaton in four-momentum $p$-space-time
is given in the form (see [11] for details)
\begin{eqnarray}
\label{e61}
\tilde W(p) = -\frac{1}{4\xi}\,i \,\frac{\partial^{2}}{\partial p^{2}}\left \{\frac{\ln\left [e^{2\gamma}(-p^{2}\,l^{2} - i\,\epsilon)\right ]}{-p^{2} - i\,\epsilon}\right \} , 
\end{eqnarray}
which leads  to the lowest order energy  of its "charge" in the static limit
$$\varepsilon (r) = i\,\int d_{3}\vec {p}\,e^{i\vec p\vec x} \,\tilde W (p^{0} =0, \vec p)\sim
\frac{1}{8\,\pi\,\xi}\, r\,\left [const + 3\ln(r/l)\right ].$$
One finds the appearance in the propagator (\ref{e61}) of the parameter $l^{-1}$ with the dimension of mass which otherwise would appear in the theory as a renormalization mass, and which distinguishes our model from the standard EW theory as conventionally formulated. 
The energy of the dilaton in the nearly conformal sector is linearly rising as $\vert \vec x\vert = r$ at large distances.

{\it Decay rate.-}
The expression for the amplitude adopted for the decay $\sigma\rightarrow \gamma\,U$
taking into account the results obtained in [12] is given by
\begin{equation}
\label{e8}
Am(x_{q},y_{q}) = \frac{3\,\alpha}{\pi\,m_{W}\,s_{W}\,\Lambda^{d-1}_{U}}
\sum_{q} c_{v}\,e_{q}  \left [I(x_{q}, y_{q}) - J(x_{q}, y_{q})\right ]
\end{equation}
with $x_{q}= 4\, m_{q}^{2}/m^{2}_{\sigma}$, $y_{q}= 4\, m_{q}^{2}/P^{2}_{U}$
for the momentum $P_{U}$ of $U$ - unparticle, the dilaton mass $m_{\sigma}$, the quarks
$q$ (in the loop) with the mass $m_{q}$ and the electric charge $e_{q}$;
$s_{W}\equiv \sin\theta_{W}$, $\theta_{W}$ is the angle of weak interactions. The axial-vector coupling $a_{v}$ does not contribute to $Am(x_{q},y_{q}) $ because of charge conjugation constraint.
We deal with the following expressions for $I(x_{q}, y_{q})$ and $J(x_{q}, y_{q})$:
$$I(x_{q}, y_{q}) =\frac{x_{q}\,y_{q}}{x_{q} - y_{q}}\left\{ \frac{1}{2} -  J(x_{q}, y_{q}) +
\frac{x_{q}}{x_{q} - y_{q}}\left [g(x_{q}) - g(y_{q})\right ]\right\}, $$
$$J(x_{q}, y_{q}) = \frac{x_{q}\,y_{q}}{2 (x_{q} - y_{q})}\left [f(y_{q}) - f(x_{q})\right ].$$

For light SM $q_{i}$ - quarks with masses $m_{q_{i}} << m_{\sigma}$ we have for $x_{q_{i}} <1$
$$f(x_{q})= -\frac{1}{4}\left (\ln\frac{y^{+}}{y^{-}} - i\pi \right )^{2}
,\,\,\,\,
g(x_{q}) = \frac{1}{2}\sqrt{1-x_{q}} \left (\ln{\frac{y^{+}}{y^{-}}} - i\pi\right ),$$
accompanied by the functions $f(y_{q})$ and $g(y_{q})$ at $1\leq y_{q} < (1-\epsilon_{\gamma})^{-1}$
\begin{equation}
\label{e10}
f(y_{q})= {\left(\sin^{-1}\sqrt {\frac{1}{y_{q}}}\right )}^{2},\,\,
g(y_{q}) = \sqrt {y_{q} - 1}\,\sin^{-1} \left (\sqrt {\frac{1}{y_{q}}}\right ),
\end{equation}
where $y^{\pm} = 1 \pm\sqrt {1 - x_{q}}$, $\epsilon_{\gamma} = 2\,E_{\gamma}/m_{\sigma}$.
The energy of the photon $E_{\gamma} = (m_{\sigma}^{2} - P_{U}^{2})/(2\,m_{\sigma})$ is restricted
in the window $[0, m_{\sigma}/2]$.

For heavy quarks   ($x_{q} \geq 1$), obeying the conditions $(4\,m_{q}/m_{\sigma}) >
x_{q} \geq  (2\,m_{q}/m_{\sigma})$, we use  (\ref{e10}) and
$$
f(x_{q})= {\left(\sin^{-1}\sqrt {\frac{1}{x_{q}}}\right )}^{2},\,\,
g(x_{q}) = \sqrt {x_{q} - 1}\,\sin^{-1} \left (\sqrt {\frac{1}{x_{q}}}\right ) $$

The energy distribution of the emitted photon in the decay width $\Gamma (\sigma\rightarrow\gamma U)$ is
$$\frac{d\Gamma (\sigma\rightarrow\gamma U)}{d E_{\gamma}} = \frac{A_{d}}{(2\,\pi)^{2}}\,m_{\sigma}\,
E^{3}_{\gamma}\left (P^{2}_{U}\right )^{d-2}\,{\vert Am(x_{q},y_{q})\vert } ^{2}, $$
where [3]
$$A_{d} = \frac{16\,\pi^{5/2}}{(2\,\pi)^{2d}}\, \frac{\Gamma(d+1/2)}{\Gamma(d-1)\,\Gamma(2d)}.$$

\begin{figure}[h!]
\renewcommand{\figurename}{Fig.}
  \centering
    \includegraphics[width=\textwidth, height = 85mm]{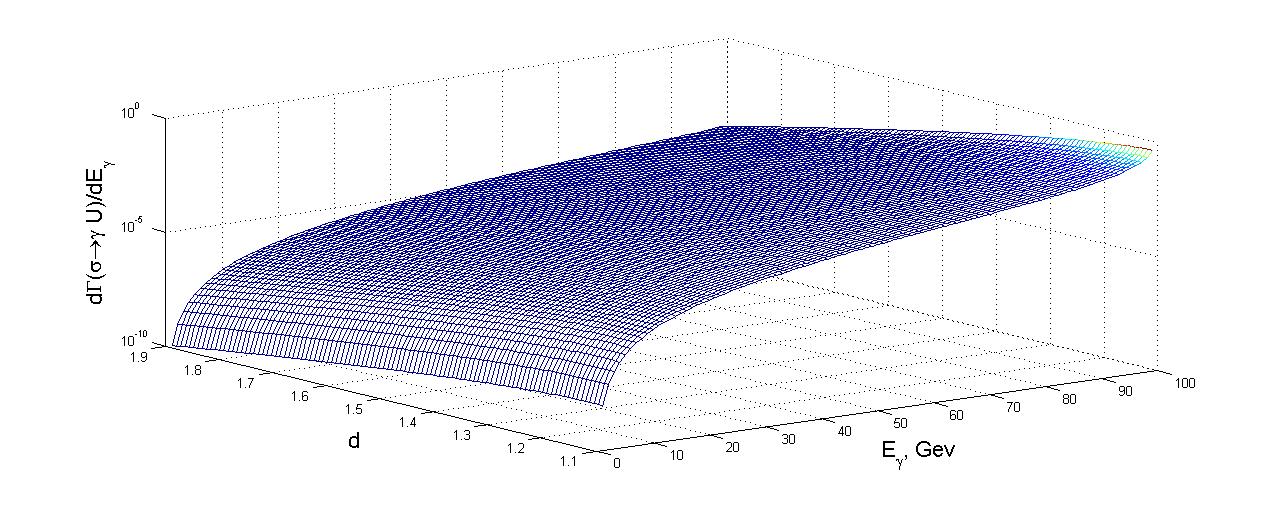}
    \caption{{ \it Energy spectrum of the photon in decay $\sigma\rightarrow\gamma U$ for various values of $d$. }}
  \label{fig:secondgraph}
\end{figure}

\begin{figure}[h!]
\renewcommand{\figurename}{Fig.}
  \centering
    \includegraphics[width=\textwidth, height = 85mm]{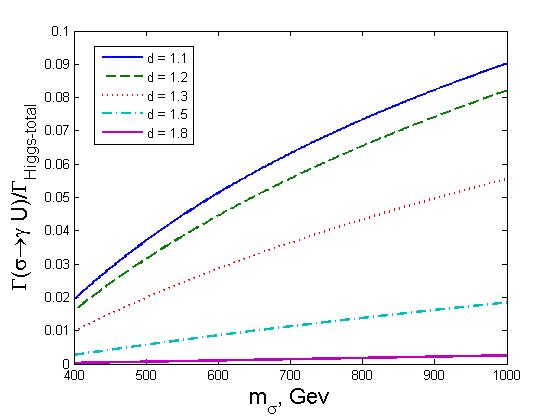}
    \caption{{ \it Decay width  $\sigma\rightarrow\gamma U$ compared to the total decay width of the Higgs-boson.  }}
  \label{fig:secondgraph}
\end{figure}

In Fig. 1, we show the energy spectrum of the emitted photon in decay $\sigma\rightarrow\gamma U$ for various
values of $d$ with the dilaton mass $m_{\sigma}$ = 200 GeV, $c_{v}$ = 1, $\Lambda_{U}$ = 1 TeV. The only top quarks in the loop are included for the calculations because of the negligible
contributions from lighter quarks ($x_{q} << 1$) in the amplitude  (\ref{e8}).

In Fig. 2, we compare the decay width $\sigma\rightarrow\gamma U$ with that of the total
decay width of the Higgs-boson. We see that this ratio increases rapidly when $m_{\sigma} \geq 400$ GeV as the scaling dimension $d$ approaches unity. The rate of the decay $\sigma\rightarrow\gamma\,U$ with respective mass range differs from that of the decay $H\rightarrow\gamma\,U$ [12] where the latter is comparable to $\gamma\gamma$- and $\gamma\,Z$- modes with the branching ratio $\sim 5\cdot 10^{-3}$ for light SM Higgs-bosom mass range up to 130 GeV.

 {\it Conclusion.-} We have studied the decay of a dilaton into a vector $U$-unparticle and a single photon. This mode is very useful to probe the nearly conformal  sector containing the dilaton and the unparticle stuff.

Unless the LHC can collect a very large sample of $\sigma$, the detection of $U$- unparticles
through $ \sigma \rightarrow \gamma\,U$ would be quite challenging.
A nontrivial  scale invariant sector of dimension $d$ may give rise to peculiar missing
energy distributions in $ \sigma \rightarrow\gamma\,U$ that can be treated in the experiment.
In particular, this energy distribution  can discriminate $d$ and estimate $\Lambda_{U}$.
When combined with $gg\rightarrow \sigma$, the decay $ \sigma \rightarrow\gamma\,U$ provides especially valuable information regarding possible loop contributions from new particles lighter than a dilaton.

The decay mode $\sigma\rightarrow\gamma\,U$ is particular useful for a heavy $\sigma$-dilaton ($m_{\sigma} > 400$ GeV) since it may have a nearly comparable decay rate with the $WW$ or $ZZ$ discovery mode of the Higgs-boson decay.

The low rate $\Gamma (\sigma\rightarrow\gamma\,U)/\Gamma_{Higgs-total}$ is compensated by the enhancement of the order $ (33/2 - n_{light})^{2}\sim O(100)$ of the gluon fusion production cross-section compared to that of the SM Higgs-boson. 

In addition to discovery, the implementation of our prediction in the LHC analyses should be straightforward and lead to more precise determination or limits of the dilaton and unparticles  couplings to gauge bosons, top quarks and the quarks of 4th generation.

\end{document}